\documentstyle[psfig]{mn}
\title[Constraints on Cosmological Models]
{Morphological Number Counts and Redshift Distributions
to {\it I} = 25 from the Hubble Deep Fields: Constraints on 
Cosmological Models from Early Type Galaxies}
\author[S. Phillipps et al.]
{S. Phillipps$^1$, S.P. Driver$^2$, Warrick J. Couch$^3$, 
A. Fernandez-Soto$^3$, \cr
P.D. Bristow$^4$, 
S.C. Odewahn$^5$, R.A. Windhorst$^6$ and K. Lanzetta$^7$\\
$^1$ Department of Physics, University of Bristol, Bristol BS8 1TL, U.K.\\
$^2$ Department of Physics \& Astronomy, University of St. Andrews, Fife, 
Scotland, U.K.\\
$^3$ Schoool of Physics, University of New South Wales, Sydney,
NSW 2052, Australia\\
$^4$ European Southern Observatory, Karl-Schwarzschild Strasse 2, D85748, Garching bei Muenchen, Germany\\
$^5$ Department of Astronomy, California Institute of Technology, Pasadena,
CA 91125, USA\\
$^6$ Department of Physics and Astronomy, Arizona State University,
Tempe, AZ 85287-1504, USA\\
$^7$ Department of Physics and Astronomy, State University of New York at 
Stony Brook, Stony Brook, NY 11794-2100, USA
}
\begin{document}
\maketitle

\begin{abstract}
We combine magnitude and photometric redshift data on galaxies in
the Hubble Deep Fields with morphological classifications in order to
separate out the distributions for
early type galaxies. The updated morphological galaxy
number counts down to $I = 25$
and the corresponding redshift distributions are used
as joint constraints on cosmological models, in particular on the values
of the density parameter $\Omega_{0}$ and normalised cosmological constant
$\lambda_{0}$.
We find that an Einstein - de Sitter universe with simple passive evolution
gives an excellent fit to the counts and redshift data at all magnitudes. 
An open, low $\Omega_{0}$, model with no net evolution (and conservation
of the number of ellipticals),
which fits the counts equally well, is somewhat
less successful, predicting slightly 
lower mean redshifts and, more significantly, 
the lack of a high--$z$ tail. A number conserving model with a dominant
contribution from $\lambda_{0}$, on the other hand,
is far less successful, predicting a much narrower distribution than seen.
More complex models are obviously possible, but we conclude that if large
scale transmutation between types does {\it not} occur, then the lambda-dominated
models provide a very poor fit to the current data.
\end{abstract}


\section{Introduction}
Galaxy number counts have long been used as potential tests of cosmological
models (Hubble 1934). However, as demonstrated by Tinsley (1972), they
are also - indeed more strongly - dependent on galaxy evolution. Once
redshift surveys at faint magnitudes became feasible (e.g., Broadhurst
et al. 1988), further constraints could be placed on the
models, breaking to some extent the degeneracy in terms of cosmological
parameters versus evolution (see, e.g., the review by Ellis 1997). More
recently, very high resolution imaging with the Hubble Space Telescope
(HST) has added a further ingredient, in that we can now identify the
morphological types of the individual very faint, distant galaxies
(e.g., Driver et al. 1995a; Glazebrook et al. 1995). This has enabled counts
to be made for the separate morphological types (e.g., Driver et al. 1995b;
Odewahn et al. 1996; Abraham et al. 1996),
removing some of the complications inherent in modelling the (varying) mix
of types in the overall counts as a function of magnitude.

In a previous paper (Driver et al. 1996, hereafter DWPB), we used number 
counts of galaxies morphologically identified as ellipticals (or S0s) in HST
Medium Deep Survey data to examine models with a wide range of the cosmological
parameters $\Omega_{0}$ and $\lambda_{0}$ (the density parameter and
normalised cosmological constant, respectively). As noted above, the 
restriction to a single type markedly reduces the complexity of the
problem since (in principle) we have only one luminosity function
and one set of k- and evolutionary corrections to deal with.
Nevertheless there remained in DWPB a range of acceptable models,
depending primarily on the amount of evolution, both stellar and dynamical,
which was allowed.

Since DWPB, numerous studies have used the yet deeper (northern)
Hubble Deep Field (HDF) data of Williams et al. (1996) to study the faint 
galaxy population (again see Ellis 1997 for a review). In the present 
paper we once more utilise morphological number counts, this time of HDF 
galaxies in both the northern and southern fields, but combine them with 
their photometric redshifts. The joint data on the galaxies
classified as early types (E/S0) are then compared with the
corresponding distributions predicted by simulations based on various
cosmologies and evolutionary schemes, in an attempt to
distinguish between the competing models. 
The simulations used, both here and
in DWPB, were described in detail in Bristow (1996; see also 
Bristow \& Phillipps 1997). 
\footnote{In brief, a volume of space is populated Monte
Carlo fashion with galaxies
following an input luminosity function. Here, as in DWPB, 
we use that of Marzke et al. (1994). In the case of the E/SOs relevant
here, they are then ascribed an intrinsic (zero redshift)
surface brightness and a corresponding scale size.
(They are all taken to have de Vaucouleurs' (1948) $r^{1/4}$ intensity
profiles). 
The corresponding observational parameters are calculated from the selected
redshift, the 
given cosmology, an input spectrum, the required bandpass and the specified
evolutionary model. (Evolution is assumed to alter only the surface brightness,
not the size, of a galaxy).
Finally, an image of the galaxy
can be created and added, along with noise etc., to a simulated data frame
with the characteristics of the actual observations used.}
 
\section{The HDF Early Type Galaxy Sample}

The basic HDF imaging data set consists of long exposures with the replacement
Wide Field Camera (WFPC2) through the four main broad band filters,
F300W, F450W, F606W and F814W. The detection of galaxies and their 
magnitude measurement were done automatically using the SExtractor 
image analysis package (Bertin \& Arnouts 1996). Details of the 
SExtractor parameters used and the zero-pointing of the `total' 
magnitudes that were measured in the analyis of the Hubble Deep Field 
North (HDF-N) data are given in Driver et al. (1998, hereafter D98); 
the analysis of the Hubble Deep Field South (HDF-S) images was done in 
an identical manner, with photometric zero-points taken from the appropriate 
STScI web-site. 

In this paper, we concentrate on the data taken through the F814W 
filter (which we will subsequently refer to as `$I$'), since it was this 
image that was used for object detection (and thus the construction of our 
galaxy catalogue) and morphological classification. The latter was 
done for all galaxies in the catalogue brighter than $I=25$, primarily 
using the ``automatic neural network" (ANN) software of Odewahn et al. 
(1996), but supplemented with visual classifications conducted by one of 
us (WJC) -- see D98 for details. Photometric redshifts, based on the 
galaxies' broad-band colours, were determined by the Stoney Brook group
in identical fashion for each sub-sample, 
and are taken from Fernandez-Soto et al. 
(1999) and Chen et al. (1998) for the HDF-N and HDF-S fields, respectively.
(The photometric redshifts are superceded by spectroscopic ones where
available).  
Further details of the combination of morphological, magnitude and 
photometric redshift data are given in D98; the galaxies classified as 
either E or S0 in the HDF-N are illustrated in the upper panel of their 
Plate L7. Note that the HDF-N observations were somewhat deeper than
those for HDF-S, hence the slightly brighter sample limit here than
in D98.

\section{E/S0 Number Counts}

HST number counts separated by morphological type have been presented by
several groups: Casertano et al. (1995), Driver et al. (1995a, 1995b),
Glazebrook et al. (1995), Odewahn et al. (1996), Abraham et al. (1996) and,
for the HDF-N, D98. The counts from these various authors show good
agreement for the early types (E/S0s), though we should note that
Marleau \& Simard (1998) have suggested that there has been a
general overestimation of their numbers due to the inclusion of round
but disc dominated systems. Our careful morphological 
examination of our sample would argue against this latter point, and in
favour of the earlier count analyses.

We showed in DWPB that the early--type 
number counts could be fitted well by (at least) three distinct 
models, differing in their cosmological parameters and assumed 
evolutionary schemes (basically allowing different amounts of dynamical 
evolution). Each of these still fits the extended data set including the 
HDF-N and HDF-S galaxies, as can be seen from Figure 1. Note that all these 
models are predicated on the assumption that the number of elliptical 
galaxies is conserved. We consider the consequences of the breakdown of 
this assumption in Section 4.

\begin{figure}
\psfig{file=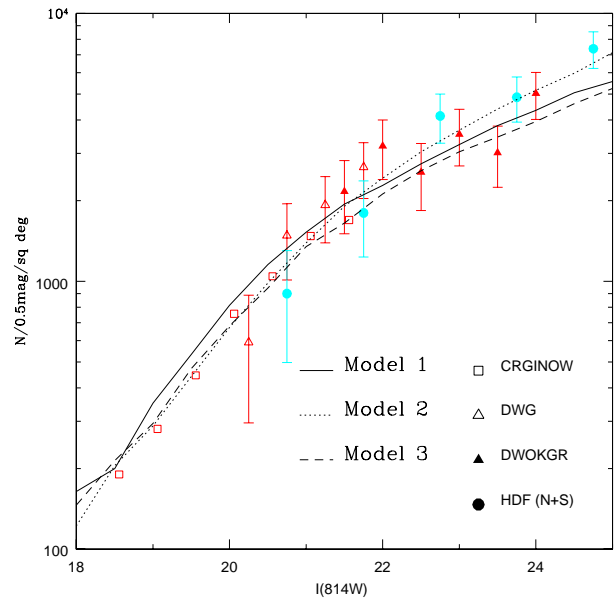,width=8.5cm}
\centering
\caption{Updated HST number counts for E/S0 galaxies. Included are the 
counts from Casertano et al. (1995 = CRGINOW), Driver et al. (1995a = DWG), 
Driver et al. (1995b = DWOKGR), plus our combined HDF-N + HDF-S counts. 
The curves represent simulated counts based on the three models 
discussed in the text.
Model 1 has $\Omega_{0} = 1$, $\lambda_{0} = 0$ and passive evolution.
Model 2 has $\Omega_{0} = 0.1$, $\lambda_{0} = 0$ and no net evolution.
Model 3 has $\Omega_{0} = 0.2$, $\lambda_{0} = 0.8$ and negative evolution.}
\end{figure}

The first successful model (number 2 in DWPB; henceforth Model 1 here)
had a standard $\Omega_{0} = 1$ Einstein - de Sitter cosmology and simple 
passive luminosity evolution approximated by $L(z) = L(0) \times (1+z)$, 
which is in fact a good approximation to the detailed predictions of
population synthesis models (e.g., those of Bruzual \& Charlot 1993,
as used by Aragon-Salamanca et al. 1998 or Roche et al. 1998) 
for evolution at red and near-infra-red wavelengths, assuming a fairly 
early formation epoch $z_{f} > 3$. (An $\Omega_{0} = 1$ model 
with no evolution underpredicts the counts). Of course, it is possible 
that the real situation is more complex than such an evolutionary 
model allows; Abraham et al.
(1999), for instance, have found evidence both for passively evolving old
ellipticals at $z \simeq 1$ and for ellipticals with recent 
star formation at $z \simeq 0.5$ in the HDF-N.

An open model,
(cf. number 3 in DWPB) with $\Omega_{0} = 0.1$ and no net evolution, 
i.e. with stellar population fading counteracted by mass evolution through 
merging, matches the counts equally well (our Model 2). Indeed, the two 
model predictions for the counts (suitably normalised) are the same to 
within 0.1 dex over the magnitude range $18\leq I\leq 25$, illustrating 
the difficulty in distinguishing between models with number counts alone 
(though the open model prediction is slightly steeper at the faintest 
magnitudes). 
McCracken et al. (2000) have similarly shown that the faint 
K-band galaxy counts (expected to be domimated by early type galaxies) 
are also well fitted by either a non-evolving low $\Omega$ model or a 
modestly evolving $\Omega = 1$ model.

DWPB also investigated various `flat' models with a non-zero cosmological
constant, i.e. models with $\Omega_{0} + \lambda_{0} = 1$. A model with 
$\lambda_{0} \simeq 0.7$, for example, is favoured by large scale 
structure arguments (e.g., Cole et al. 1997), and by the comparison of 
cosmic microwave background and SNIa data (Lineweaver 1998).  For a given
evolutionary scheme, increasing $\lambda_{0}$ at the expense of $\Omega_{0}$
significantly steepens the predicted counts, so in order to make such a 
model (with conserved numbers) fit the data it is necessary to adopt 
extreme negative evolution, i.e. high--$z$ ellipticals were so much smaller 
(less massive)
than today's as to overwhelm the effect of their stellar population 
brightening. This may not be too unreasonable at high $z \sim 1 - 2$ (cf. 
D98). Our Model 3 thus has $\Omega _{0} = 0.2$, $\lambda_{0} = 0.8$ and 
$L(z) = L(0) (1+z)^{-1}$, which again matches the counts and is 
essentially indistinguishable from the other two models
across our observed magnitude range, again emphasising the need for a
`tie-breaker'.

\section{E/S0 redshift distribution}
As the additional constraint which allows us to discriminate between the 
different models, we take the combined redshift distribution of E/S0 
galaxies in the two Hubble Deep Fields. By combining the HDF-N and HDF-S 
data sets, we improve the statistics and reduce the effects of clustering 
along any one line of sight. Furthermore, comparisons with spectroscopic 
redshifts have shown that photometric redshifts are generally accurate 
to 0.1 or better (Fernandez-Soto et al. 1999), which is certainly more 
than adequate for our purpose. 

First consider the brighter galaxies in our sample (for which there is least ambiguity in morphology).
Figure 2 shows the E/S0 redshift distribution observed in the range 
$22\leq I\leq 24$, along with the predictions of the three models,
all normalised to the observed number of galaxies. 
Compared to the open model, luminosity distance $D_{L}$ is smaller at 
given $z$ for the $\Omega_{0} = 1$ model, so conversely $z$ is larger at 
given $D_{L}$ and hence at any given distance modulus or apparent 
magnitude. Since the model fit to the counts requires net positive 
luminosity evolution (galaxies brighter at larger $z$) in the $\Omega_{0} 
= 1$ model (but not in the open model), this further increases the mean 
$z$ at any magnitude. The greater difference between the models at larger 
$z$ also tends to spread out the expected $z$ distribution for 
$\Omega_{0} = 1$, leading to a significant high--$z$ tail.

The general distribution of redshifts seen in Figure 2 is, in fact, in
remarkably good agreement with that predicted for the Einstein - de Sitter
plus passive evolution model (left hand panel). 
We do not have sufficient numbers to test the
shape of the distributions in any detailed way, but we can see that
at $22 < I < 24$, the mean redshift $<$$z$$>$ $\simeq 0.75$ compared
to the model prediction of 0.76. Furthermore, 6 galaxies ($15\pm 5$\%) have
redshifts in the range $1.0\leq z\leq 1.5$ compared to a predicted 21\%. 
The non-evolving $\Omega_{0} = 0.1$ model (centre panel)
has a predicted $<$$z$$>$ = 0.65, 
i.e. lower, but probably not significantly so. However,  only 4\% of 
objects (i.e. 1--2) should be at $z > 1$, in substantial conflict with the 
data here and in D98. If this model were correct we would expect to see
many more of the galaxies at $0.4 < z < 0.8$.

In the $\Lambda$--dominated models, distances are even larger at given
$z$ than for the open model, so basically we do not expect to see
to very large $z$. This is compounded by the fact that we need
negative evolution to fit the counts in this model. The predicted
maximum $z$ is therefore lower at each apparent magnitude. In
addition we expect the distribution to be narrowed on the other side 
because of the relatively smaller volume at low $z$. 
Thus, although $<$$z$$>$ is similar to that in the other models (0.71), there 
should be essentially no galaxies seen at $z > 1$ or $z < 0.4$. 
{\it Both} these aspects make the high $\lambda_{0}$ model 
(right hand panel of Figure 2) a very poor fit to 
our observed $n(z)$.

At fainter magnitudes,
the predicted mean $z$ continues to grow
substantially for the E-deS model (less so for the open  and lambda models) and
tracks the observed mean $z$ really rather well; $<$$z$$>$ = 1.01 predicted
at $24 < I < 25$ compared to $<$$z$$>$ $\simeq 1.05$ observed in the present
sample and 1.25 predicted, 
$\simeq 1.3$ observed at $25 < I <26$ for the HDF-N (see D98: 
this excludes the handful of 
objects apparently at $z > 3$ in D98). The general wide spread of redshifts
in each magnitude slice (D98)
is clearly more characteristic of the evolving 
$\Omega_{0} = 1$ model, too.

\begin{figure*}
\vspace*{-8.5cm}
\psfig{file=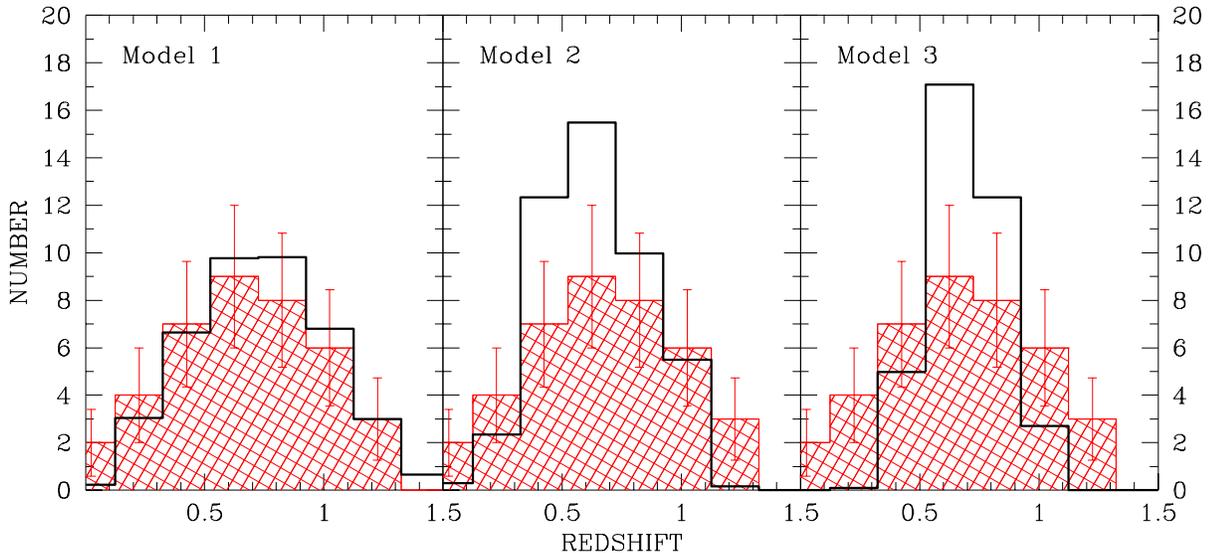,width=17.0cm}
\vspace{-0.9cm}
\centering
\caption{
The distribution of photometric redshifts for HDF-N and HDF-S E/S0 galaxies 
with $22 < I < 24$ (cross--hatched histogram and poisson error bars) 
compared to the predictions of each of
the three models shown in Figure 1 (solid lines). All the 
model histograms are 
normalised to the same total number of galaxies as in the data 
histogram.} 
\end{figure*}

\section{Discussion and Summary}

From a theoretical galaxy formation point of view, one might nowadays 
perhaps favour the non-evolving, open universe fit to the
counts, since dynamical evolution via mergers (e.g., Kauffmann \& White 
1993) {\it is} expected to counteract quite closely the fading of stellar 
populations with time (see DWPB). Indeed, in the context of brightest 
cluster galaxies, Aragon-Salamanca et al. (1998) have found that their 
empirical result of {\it no net} evolution in luminosity of BCGs for 
$\Omega_{0} = 0$ requires an amount of growth by mergers which is matched 
almost exactly by current semi-analytic models in open CDM universes
(in particular, those of Baugh et al. 1996
and Kauffmann \& Charlot 1998). Thus, in this model, both the BCG
and general E/S0 galaxy dynamical evolution are the same and matched
by theory. (Further, Le Fevre et al. (2000) have found a variation
of galaxy close pair fraction with $z$ which would be consistent
with mergers occuring at the correct rate to more or less cancel
the stellar evolution). 

In order to make this model fit our redshift data, we would have to explain 
away the high--$z$ tail of the distribution. (The apparent {\it lack} of
a high--$z$ tail in other data, for instance moderately deep K-band 
selected samples, has sometimes been used as evidence {\it for} the 
model where mergers induce zero net evolution, see, e.g., Kauffmann \& 
Charlot 1998). Accordingly, we have checked the credentials of objects that 
populate the high--redshift tail of Figure 2
to eliminate the possibility that they 
are in any way spurious. Firstly, we conducted a visual inspection of 
each object to make sure that this subset of the population is not 
contaminated by misclassified early--type spirals nor any other type of 
object. In all cases the classification of these objects as E, E/S0, S0 or 
S0/a types was categorical. A second possibility is that the redshifts 
of these objects have been overestimated. However, 5 of the 6 galaxies 
have now had their redshifts measured spectroscopically and are thus quite 
secure. Hence we see no reason to doubt that the population of E/S0s at 
$z>1$ is both real and significant in number. 

For $\Omega_{0} = 1$ universes, as favoured by our
redshift results, Aragon-Salamanca et al. (1998)
find that the BCGs must have undergone {\it negative} net evolution,
i.e. stronger merging,
as indeed is appropriate to the semi-analytic models for $\Omega_{0} = 1$ CDM.
This is the opposite of what we would require for the counts in
this case. In other words, for the $\Omega_{0} = 1$ model to fit
both the BCG luminosities {\it and} the general E/S0 galaxy counts we
need strong merger evolution of the BCGs but essentially none
(i.e. passive evolution only) for early type galaxies as a whole.
Interestingly, Mobasher \& Trentham (1998) have recently shown, on
the basis of $K$-band luminosity functions of galaxy clusters, that 
the latter, at least, does appear to be the case. They find no evidence
for an increase in the stellar mass of ordinary
early type galaxies since $z = 0.9$,
the data being consistent with purely passive evolution.
Kodama, Bower \& Bell (1999) have also found that the colour --
magnitude relation for early type galaxies at $z \sim 0.9$ (specifically HDF
galaxies as studied here) is consistent with passive evolution from
that epoch until the present (though the shallow slope
of the C--M relation makes it a weak discriminant of mass evolution). (See also
Moles et al. 1998 for other arguments against significant dynamical
evolution at recent epochs). Blain et al. (1999) also tentatively support
an $\Omega = 1$, rather than an open or lambda--dominated model to
explain the sub-mm counts and infra-red background data. This would be
in line with our current result. Note that a prediction of this model,
assuming the BCG result to be correct, is that $K_{\star} - K_{BCG}$
should be smaller at higher $z$. Of course, it is possible that the
true situation

In a lambda--dominated model, distances to
a given $z$ are much higher than in the conventional models, so
the Aragon-Salamanca et al. BCGs would have to have strong {\it positive}
evolution, i.e. a net brightening in the past. The difference between the
required evolution of BCGs and ordinary ellipticals noted above for
the $\Omega_{0} = 1$ model would then be reversed, little or none for BCGs and large
amounts for ellipticals in general. This might well be considered a
much less palatable option. Note that
the conflicting requirements for
the counts and the BCG Hubble diagram,
here (and in the $\Omega_{0} = 1$ case), arise because while
distances are greater than in the open model, making galaxies 
look fainter, there is more volume, thus steepening 
the counts (the opposite ocurring for $\Omega_{0} = 1$). Avoiding the
requirement for large negative
evolution would imply a very significant undercounting of ellipticals
in the HDF data. (The opposite of the effect claimed in the one non-concordant
analysis (Marleau \& Simard 1998).)
Even a model with no evolution predicts twice as 
many galaxies at $I = 25$ as the successful models (cf. DWPD).

In addition to any problem with the evolution, making
the lambda model which fits the counts also fit our $n(z)$ requires us to have 
mistakenly included non-ellipticals at both high and low $z$. This would
clearly make the real counts lower, making it even harder for the 
lambda--dominated model to fit them. (Alternatively we might have missed a 
number of medium $z$ ellipticals, so that the real counts should be 
higher and the $z$ distribution more peaked, but it is hard to see how this
might occur). These problems are in apparent conflict with the
recent arguments (e.g., Lineweaver 1998) in favour of
models with a significant cosmological constant. We discuss this further below.

The three models from DWPB which fit the counts also represent the
three generic types of model currently favoured by large scale
structure arguments (e.g. Lin et al. 1996; Cole et al. 1997). 
The three most studied ways out of the problems faced
by the standard cold dark matter picture (SCDM) are to keep $\Omega_{0} = 1$
but change the initial fluctuation spectrum (tilted CDM), move to a low
$\Omega$ universe (open CDM), or invoke a non-zero cosmological constant
(lambda CDM). 
\footnote{Note that this is only a restricted identification of our
empirical models with the CDM models; it does not (necessarily) imply
the whole structure formation picture based on the CDM variants,
only the use of the same values of the cosmological parameters.}
Thus the attempt to break the degeneracy of these models
in the galaxy counts also has relevance to the large scale structure
problem. If $\Omega_{0}$ really is unity, as suggested by
the counts and redshifts, 
this might indicate that it is the primordial
spectrum which needs to be adjusted. A high $\Omega_{0}$ is
consistent with data on Cosmic Microwave Background
fluctuations (Turner 1999), which primarily
constrain $\Omega_{0} + \lambda_{0}$ to be near unity
(de Bernardis et al. 2000), and with supernova
observations at moderate redshift (Perlmutter et al. 1997). However, it
is incompatible with the Hubble diagram for high--redshift supernovae
(Perlmutter et al. 1998, Reiss et al. 1998). The $\Omega_{0} = 1$ model
also has problems with, for instance, the observed high baryon fraction
in clusters, and generally scores poorly in Peebles' (1999) summary of current
tests of viable world models. 

To summarise, then, following up the work of DWPB, who found a range of
possible 
cosmological models which could fit the number counts of early type
galaxies observed by HST, we have addded a further constraint in
the distribution of (photometric) redshifts for very faint 
(northern and southern) HDF
E/S0 galaxies.  We find remarkably good agreement in the case of
a standard Einstein - de Sitter universe with passively evolving
early type galaxies. If the preference
for a high density model {\it is} confirmed (and see above for
arguments against it), then it appears that we require
significantly different evolution for ellipticals as a whole (i.e.
negligible merging at recent epochs) compared to first ranked cluster 
galaxies used in the Hubble diagram (which are consistent with the 
theoretically expected amount
of merging more or less cancelling the stellar population evolution).

The perhaps initially preferred open model
with no net evolution (i.e. stellar fading cancelled by growth
through mergers) proves to be somewhat less successful, primarily because it
does not predict sufficient galaxies in the high--redshift tail. 
Further confirmation that these objects really are genuine ellipticals at
$z > 1$ would obviously be of great value here.
Even larger problems face the currently favoured models dominated 
by a cosmological constant (Lineweaver 1998, Turner 1999, Peebles 1999).
Those which can match the number
counts predict a narrow range of redshifts ($\simeq 0.5 - 1$), 
quite unlike the fairly flat distribution out to high--$z$
actually seen. They also appear to
require the unlikely circumstance of no merger
evolution for BCGs but large amounts (very small precursors at
moderate $z$) for ellipticals in general.

We have already considered conceivable problems with the data which might 
allow consistency with the low $\Omega$ models. 
On the modelling side, the faint end
of the E/S0 luminosity remains problematic (see DWPB).
Possibly more importantly, our models all conserve the number of large
(i.e. potentially visible) systems, i.e. there is no number evolution 
(see He \& Zhang 1999). Even though we allow merger evolution,
we do not specifically increase the normalization of the LF with $z$
(unlike, say, Rocca-Volmerange \& Guiderdoni (1990) who increase their
Schechter function normalisation in step with the decrease in 
characteristic luminosity). This means that we are essentially
allowing largish ellipticals to grow further by accreting
small neighbours, rather than by equal mass mergers. At the
opposite extreme, we can simply double the numbers of visible galaxies at
$z \sim 1$ for the $\Omega_{0} = 1$ model, or multiply by 4 for
the lambda model (consistent with the required changes in $L$ via
mergers, in each case). Even this predicts only 8\% or 1\% of galaxies in 
the high--$z$ tail, still far fewer than actually seen.
Finally, none of our models allow for the wholesale transmutation 
between galaxy types. 
If there were {\it fewer} early type galaxies
at higher $z$, that is they continue to {\it form}
(as opposed to grow) recently via mergers,
fading or harrassment (Moore et al. 1998 and references therein), this
would not change our conclusions since an even smaller
high--$z$ tail would be predicted in models which already underpredict
the numbers seen at large $z$. On the other hand,
the existence, temporarily, of additional
spheroidal merger remnants which then grow new discs 
(Windhorst et al. 1998a,b; Scoville et al. 1997;
Pascarelle et al. 1996) might perhaps allow
a better fit for the low $\Omega$ models, the observed high--$z$ objects
then being from effectively a separate population. We will consider 
whether any 
possible mechanisms of this sort might allow consistency of the
counts and redshift data with the currently popular lambda--dominated models
in a subsequent paper. For now, we conclude that if large scale
transmutation of types does not occur, then the lambda--dominated
models are a very poor fit to the current data.

\section*{Acknowledgements}
We would like to thank Richard Bower, Malcolm Bremer, Matthew Colless,
Roberto de Propris, Carlos Frenk,
Bryn Jones, Charlie Lineweaver, Henry McCracken, Brian Schmidt and
Tom Shanks for useful comments. SP was supported by the Royal Society
via a University Research Fellowship during the course of this work. 
WJC, SPD and AFS acknowledge the 
financial support of the Australian Research Council and the Department 
of Industry, Science and Technology at various stages of this project.

\end{document}